# Electrokinetically-driven deterministic lateral displacement for particle separation in microfluidic devices

Srinivas Hanasoge, Raghavendra Devendra, Francisco J. Diez and German Drazer

Mechanical and Aerospace Engineering Department,

Rutgers, The State University of New Jersey,

Piscataway, NJ.

**Abstract**

An electrokinetically-driven deterministic lateral displacement (e-DLD) device is proposed for the continuous, two-dimensional fractionation of suspensions in microfluidic platforms. The suspended species are driven through an array of regularly spaced cylindrical posts by applying an electric field across the device. We explore the entire range of orientations of the driving field with respect to the array of obstacles and show that, at specific forcing-angles, particles of different size migrate in different directions, thus enabling continuous, two-dimensional separation. We discuss a number of features observed in the kinetics of the particles, including directional locking and sharp transitions between migration angles upon variations in the direction of the force, that are advantageous for high-resolution two-dimensional separation. A simple model based on individual particle-obstacle interactions accurately describes the migration angle of the particles depending on the orientation of the driving field, and can be used to re-configure driving field depending on the composition of the samples.

INTRODUCTION

A number of microfluidic devices have been proposed in recent years for the continuous separation of suspended particles employing novel techniques.[1] A popular separation method is deterministic lateral displacement (DLD), in which a mixture of species is driven through an array of cylindrical posts. The interaction of the different species with the posts leads to different species migrating in different directions with respect to the array, thus causing the two-dimensional and continuous fractionation of the mixture.[2,3] Microfluidic DLD devices have been successfully applied to the fractionation of different mixtures, and in particular for cell sorting and the separation of biological material.[4] In recent work, we demonstrated a unique operating mode for DLD separation based on a uniform force driving the suspended particles, in particular, gravity (g-DLD).[5,6] In terms of static (no-flow) systems, a powerful and ubiquitous method to affect separation, in both traditional as well as microfluidic systems, is to drive the suspended mixtures using electric fields. Surprisingly, in spite of its potential, the use of electric fields in DLD systems has not been investigated, possibly due to the fact that the original explanation of the separation phenomena observed in DLD systems was based on the flow field.[2,3] However, in previous work, we have shown in experiments and simulations, that particle separation in DLD systems is induced by the cumulative and *separative* effect of particle-obstacle non-hydrodynamic interactions, whose presence is independent of the driving field.[7–11]

Here, we demonstrate the simplicity and potential of electrokinetically-driven DLD (e-DLD) separation devices using a simple microfluidic system that can be easily re-configured to control the relative orientation of the driving field with respect to the array of posts. We performed detailed experiments tracking the motion of individual particles of three different sizes for the entire range of possible orientations of the driving electric field. We show that the kinetics of individual particles is completely analogous to that observed in flow- and force-driven DLD systems. More important, we identify specific angles at which different particles migrate in different directions within the array. Moreover, we show a significant difference in the migration angles, providing excellent size resolution and demonstrating the potential of e-DLD as a two-dimensional and continuous separation method. In addition to extending the versatility of DLD with an alternative driving field, the use of electric fields opens the possibility to on-line control of the orientation of the driving field depending on the sample.

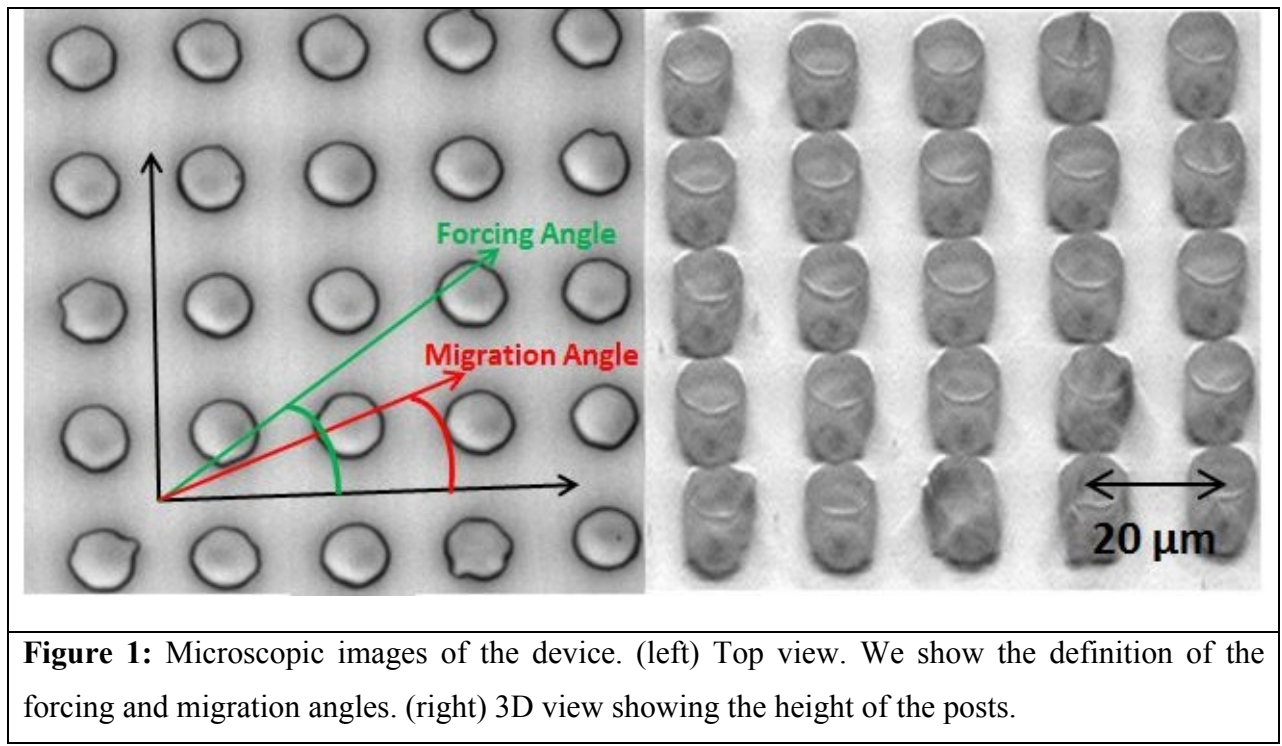

**Figure 1:** Microscopic images of the device. (left) Top view. We show the definition of the forcing and migration angles. (right) 3D view showing the height of the posts.

**EXPERIMENTAL**

**Materials and Fabrication:**

The particles used in the experiments are: 4.32 $\mu m$ (silica, density 2 $g/cm^3$, Bangs Laboratories, Inc., CA), 10 $\mu m$ and 15 $\mu m$ (glass, density 2.5 $g/cm^3$, Duke Scientific, CA). The particles were suspended in a 1% aqueous solution of borate buffer to maintain a constant pH during the experiments. The array of cylindrical posts was fabricated in a negative photoresist (SU8 2025 - Microchem Corp., MA) directly on a microscope glass slide using standard photolithography process. A picture of the device is shown in Figure 1. The micro-posts were approximately 19 $\mu m$ in diameter, 35 $\mu m$ in height and separated by 40 $\mu m$ center-to-center. A channel fabricated in polydimethyl siloxane (PDMS) using softlitography methods was then used to cover the array of posts as indicated in Figure 2. The dimensions of the channel were 50 $\mu m$ in depth, 3 mm wide, and 2 cm long. Although the channel was higher than the obstacles, the particles deposit to the bottom of the channel and move around the obstacles. Two relatively large reservoirs (1cm × 0.5 cm, see Figure 2) are punched into opposite ends of the PDMS channel to reduce the build-up of pressure during the experiments.[12]

**Experimental setup and Method:**

The applied electric field across the PDMS microchannel generates the electroosmotic flow that drives the particles. In addition, the particles experience an electrophoretic force that depends on their charge, but their motion is usually dominated by the electroosmotic flow, specially in buffered solutions.[13–16]

The direction of the driving field is controlled by the orientation of the PDMS channel with respect to the array of obstacles. Therefore, we performed experiments for different angles of the driving field by changing the orientation of the channel, which was reversibly attached to the glass slide. The fact that the channel determines the direction of the driving field was verified by tracking the motion of particles inside the channel but outside the array of posts. Specifically, at the beginning of each experiment, the forcing angle relative to the obstacle array, θ, is determined experimentally by tracking the motion of particles in an obstacle-free region inside the microchannel. In order to avoid spurious correlations in the results, the experiments are not carried out in any particular decreasing or increasing order of forcing angles. The migration angle relative to the obstacle array, α, is then measured by tracking an independent set of particles moving through the array of posts. In each of the experiments, we tracked around 15-20 different particles of the same size, both in the obstacle-free region as well as inside the array. In order to reduce fluctuations in the measured migration angle, only those trajectories that are 100 µm or longer - in the horizontal direction – are considered. In addition, we only considered the motion of particles that did not interact with other particles along their trajectories. The definition of the migration and forcing angles (as well as the coordinate axes) are illustrated in Figure 1.

The experiments are performed in the deterministic limit, that is, when the diffusive transport of the suspended particles is negligible. As an estimation, the Peclet numbers, calculated from the Stokes-Einstein relation for the bulk diffusivity of the particles and the experimentally measured average velocity, range from $O(10^2)$ to $O(10^4)$.

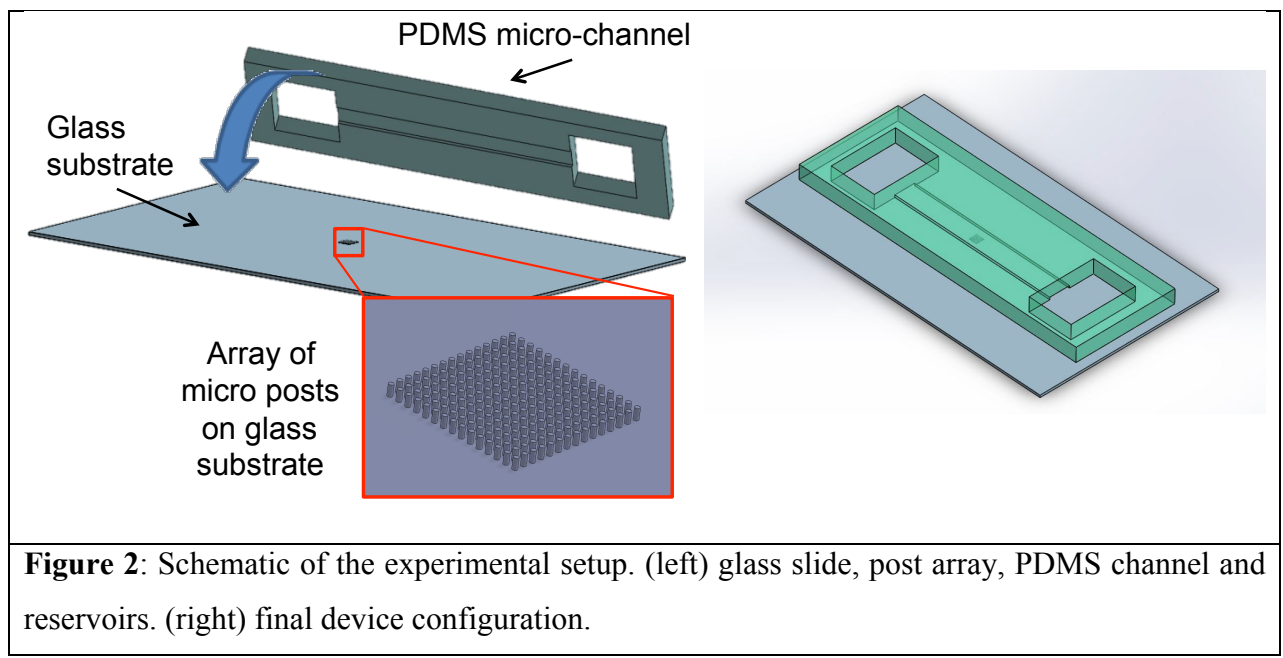

**Figure 2**: Schematic of the experimental setup. (left) glass slide, post array, PDMS channel and reservoirs. (right) final device configuration.

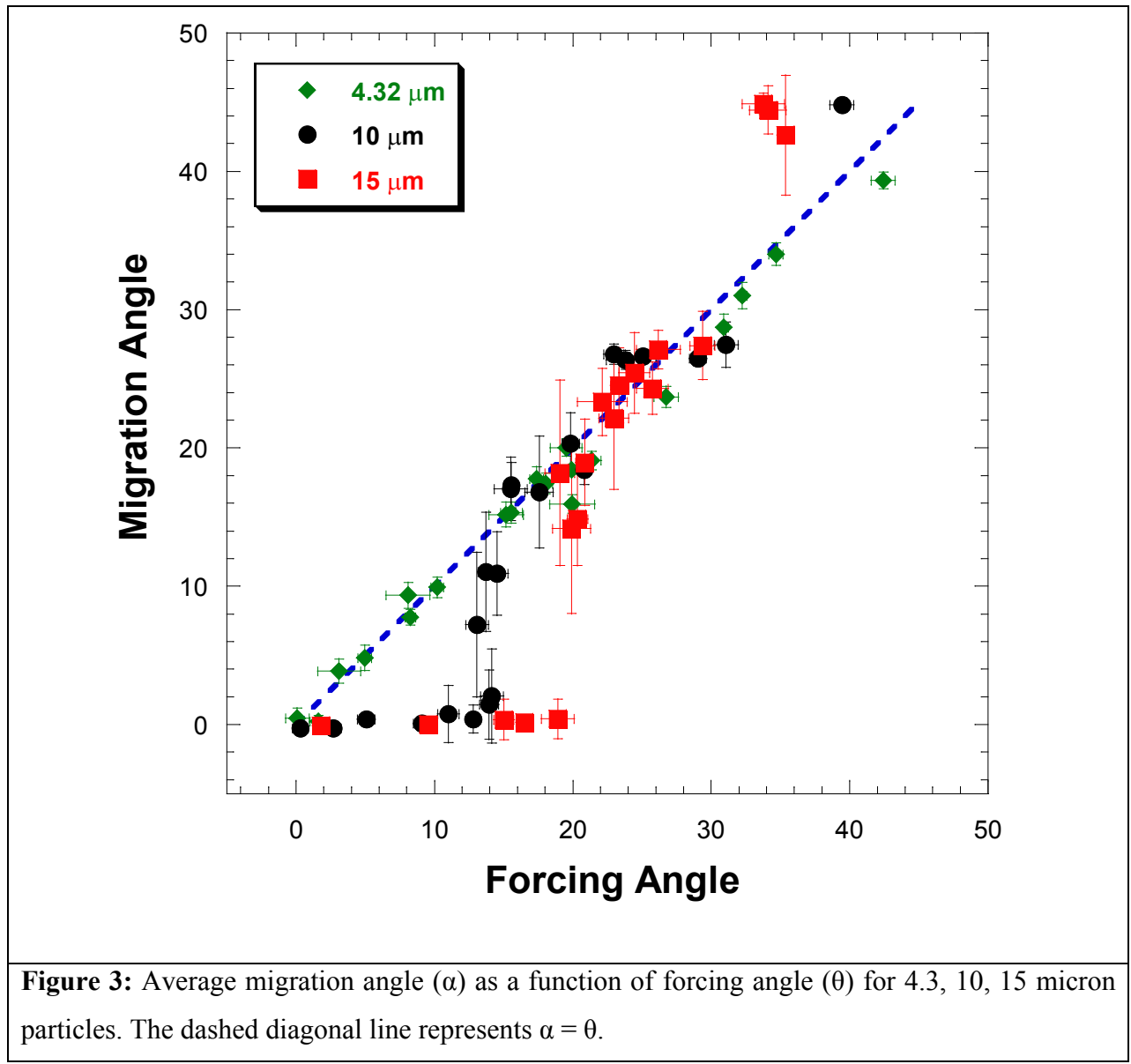

**Figure 3:** Average migration angle (α) as a function of forcing angle (θ) for 4.3, 10, 15 micron particles. The dashed diagonal line represents $\alpha = \theta$.

**RESULTS AND DISCUSSION**

We have performed a series of independent experiments to measure the migration angle of suspended particles of sizes 4.32 $\mu m$, 10 $\mu m$ and 15 $\mu m$ electrokinetically-driven through the obstacle array at forcing angles, θ, ranging from 0° to 45°, with respect to the main lattice directions (see Figure 1). In Figure 3, we present the migration angle, α, versus forcing angle, θ, for all particles over the complete range of forcing directions. (Note that each data point corresponds to independent experiments, in which the average migration angle is determined for a given forcing orientation). The vertical (horizontal) error bars represent the standard deviation in the measured migration (forcing) angles. The observed motion of the suspended particles is completely analogous to that observed in previous DLD experiments, with particles exhibiting periodic trajectories and directional locking. That is, particles move at certain lattice directions that remain constant for a range of forcing angles. For example, all particles are *locked* in the [1,0] direction at small forcing angles. Essentially, particles move along a line of consecutive posts in one of the principal directions of the array (say a *column* of the array). Only at large enough forcing angles the particle move across a column of obstacles in the array and their migration angle becomes different from [1,0]. We also observe sharp transitions between locking directions, a phenomena that could be harnessed to obtain high selectivity and resolution in the separation process.[5] It is also clear that, at certain special forcing angles, particles of different size exhibit different migration angles, which is the basis for separation. Interestingly, the first transition angle (or *critical angle*) $\theta_c$, defined as the largest forcing angle for which the particles are locked to move in the [1,0] direction (along a column of the array), shows the largest variation with the size of the particles, with smaller particles transitioning at smaller angles. More specifically, the smallest particles, 4.32 μm, exhibit a very early transition at $\theta \sim 5^0$, while 10 μm and 15 μm particles remain locked to move in the [1,0] lattice direction until about $\theta \sim 14^0$ and $\theta \sim 19^0$, respectively. Therefore, a driving direction $\theta \sim 18^0$ would efficiently separate the 4.32 μm particles from the 15μm ones. In fact, in Figure 4, we show two sets of trajectories obtained from 4.32 μm and 15μm particles at $\theta \sim 18^0$, and a significant difference in migration angles is evident.

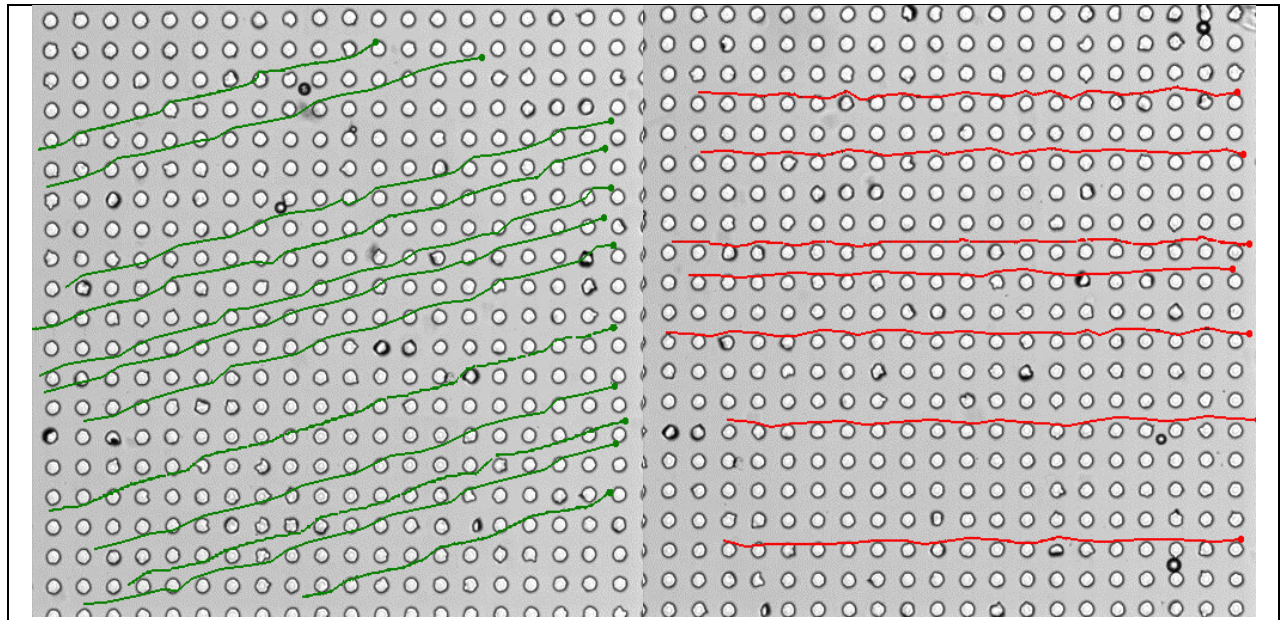

**Figure 4:** Trajectories obtained for a forcing angle $\theta \sim 18^0$ with (left) 4.32 μm and (right) 15 μm particles. (left) Smaller particles are able to closely follow the forcing angle, as shown in Figure 3. (right) Larger particles are locked into the [1,0] lattice direction. As a result, there is a significant difference in migration angles, nearly as large as the forcing direction.

Figures 5 shows the average migration angles obtained in each of the individual species. The figures also show the theoretical predictions based on a simple *collision model* developed to explain g-DLD experiments.[6,7] In this simple model, particle-obstacle collisions might induce a net lateral displacement in the trajectory of the suspended particles, depending on the initial offset of the collision (defined as the distance between the incoming particle and a line in the direction of the driving force that goes through the center of the obstacle). Specifically, for any initial offset $b$ smaller than a critical offset $b_c$, the offset after the collisions becomes equal to $b_c$. It is this irreversible collapse of trajectories that leads to the observed directional locking.[8] The value of $b_c$ for each particle size is obtained by fitting the experimental data and is indicated in the figures, and is related to the first critical angle by $b_c = l \sin(\theta_c)$, where $l$ is the post-to-post distance (the separation between posts in the square array). We observe that the value of the critical offset $b_c$ increases with size of the particles, which corresponds to the increase in the first critical angle discussed before. Overall, there is excellent agreement between the model and the experimental results for all particle sizes. Note that, $b_c$ is the only fitting parameter, and for a given value of $b_c$, the model predicts all the possible locking directions as well as the forcing angles at which there is a transition in the locking directions.

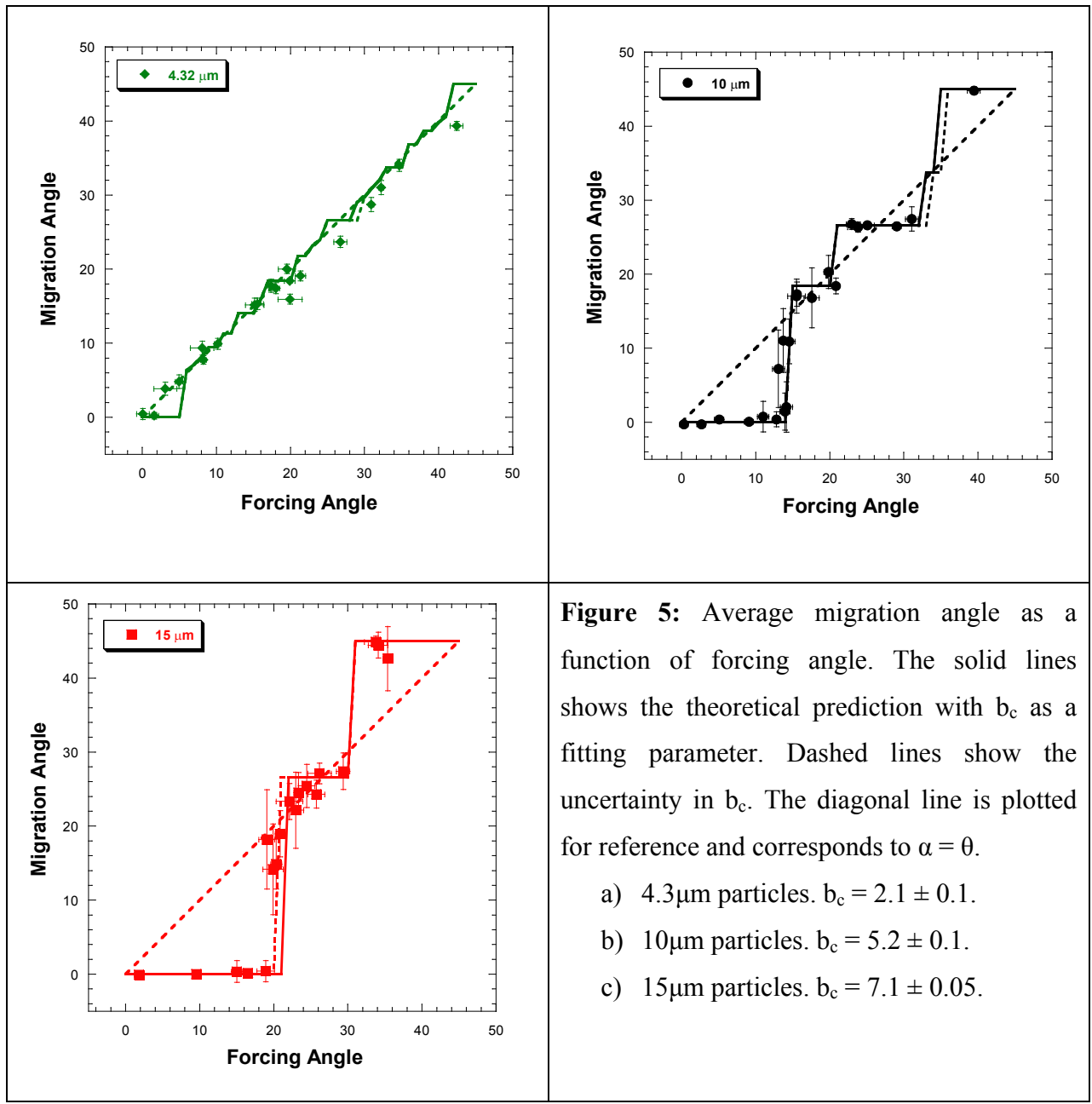

**Figure 5:** Average migration angle as a function of forcing angle. The solid lines shows the theoretical prediction with $b_c$ as a fitting parameter. Dashed lines show the uncertainty in $b_c$. The diagonal line is plotted for reference and corresponds to $\alpha = \theta$.

a) 4.3μm particles. $b_c = 2.1 \pm 0.1$.
b) 10μm particles. $b_c = 5.2 \pm 0.1$.
c) 15μm particles. $b_c = 7.1 \pm 0.05$.

**CONCLUSIONS**

We have demonstrated the basis for electrokinetically-driven deterministic lateral displacement (e-DLD) for size-based separation of suspended particles in microfluidic devices. We have shown that, the observed deterministic kinetics of the particles is analogous to the case of flow- and force-driven DLD, including directional locking, and our understanding of those systems can be applied to describe and analyze the behavior in the proposed e-DLD devices. We performed a comprehensive set experiments that covered the entire range of orientations of the driving electric field, and showed the potential for separation at specific forcing angles. We also showed that the first transition angle exhibits a large dependence on particle size, with a difference of nearly $15^0$ between the smallest and largest particles, which suggests the use of relatively small angles to optimize the resolution of the system. The differences in the critical angles, the significant difference in the angle of the observed locking directions ($\approx 18^0$ between [1,0] and [1,3] directions, or $\approx 26^0$ between [1,0] and [1,2] directions), and the sharp transitions between locking directions, indicate the potential for high size resolution of the proposed e-DLD separation method.

The use of electrokinetically-driven flows could significantly expand the application of DLD methods. Driving the suspension with electric field is also amenable of on-line external control and re-configuration of the forcing angle depending on the sample, something that is not straightforward in flow-driven DLD. Finally, let us mention that the use of electric-fields provides ways to control or tune the operation of DLD devices, by manipulating not only particle-obstacle interactions, as shown in recent work[17], but also the spatial variations of the driving field to enhance the separative displacement among different species.

This work is partially supported by the National Science Foundation Grant No. CBET-1343924.